\documentclass[useAMS,usenatbib]{mn2e}
\pdfoutput=1

\usepackage{graphicx}

\def\EBV{\mbox{$E(B-V)$}}

\def\mMo{\mbox{$(m-M)_o$}}
\def\ms{\mbox{$M_\odot$}}
\def\zs{\mbox{$Z_\odot$}}
\def\ds{\mbox{$d_\odot$}}
\def\fB{\mbox{$f_{bin}$}}
\def\fI{\mbox{$f_{ind}$}}
\def\my{\rm Myr}
\def\gy{\rm Gyr}

\title[Isochrone fits to CMDs]{Extracting parameters from Colour-Magnitude Diagrams}

\author[C. Bonatto, F. Campos, S.O. Kepler, and E. Bica]
{C. Bonatto$^1$, F. Campos$^1$, S.O. Kepler$^{1}$ and E. Bica$^{1}$\\
$^1$ Departamento de Astronomia, Universidade Federal do Rio Grande do Sul, 
Av. Bento Gon\c{c}alves 9500\\ 
Porto Alegre 91501-970, RS, Brazil}

\begin{document}

\pagerange{\pageref{firstpage}--\pageref{lastpage}}

\maketitle

\label{firstpage}

\begin{abstract}
We present a simple approach for obtaining robust values of astrophysical parameters from the 
observed colour-magnitude diagrams (CMDs) of star clusters. The basic inputs are the Hess diagram 
built with the photometric measurements of a star cluster and a set of isochrones covering wide 
ranges of age and metallicity. In short, each isochrone is shifted in apparent distance modulus 
and colour excess until it crosses over the maximum possible Hess density. Repeating this 
step for all available isochrones leads to the construction of the solution map, in which the 
optimum values of age and metallicity - as well as foreground/background reddening and distance 
from the Sun - can be searched for. Controlled tests with simulated CMDs show that the approach 
is efficient in recovering the input values. We apply the approach to the open clusters M\,67,
NGC\,6791, and NGC\,2635, which are characterised by different ages, metallicities and distances 
from the Sun.
\end{abstract}


\section{Introduction}
\label{intro}

Colour-magnitude diagrams (CMDs) encapsulate important properties of single or composite stellar 
populations. The most relevant are the age, metallicity, binary fraction and, perhaps, the initial 
mass function. These are followed by the foreground/background reddening, distance from the Sun, 
differential reddening (especially for embedded clusters), and even age spread (for young clusters) 
of the constituent stars. Combined, these parameters contribute - in different degrees - to shaping
and defining the CMD morphology of a star cluster. Obviously, the definition of the evolutionary 
sequences of a single stellar population in a CMD increases with the number of member stars.

A proper or robust determination of astrophysical parameters of star clusters is an important 
source of constraints for tracing the Galactic structure and understanding its dynamics. 
Fundamental parameters of young star clusters are important also for studies of dynamical state 
and cluster dissolution time scales (e.g. \citealt{GoodW09}; \citealt{Lamers10}), {\em infant
mortality\footnote{Essentially, the early ($10\sim40\,\my$) phase in cluster evolution affected
by significant disruption processes.}} (e.g. \citealt{LL2003}; \citealt{GoBa06}), star-formation 
rate in the Galaxy (e.g. \citealt{LG06}; \citealt{SFR}), among others. This kind of analysis 
started with \citet{Trumpler30} and continues to date (e.g. \citealt{CBB13} and references therein).

In this context, the existence of methods to help in the extraction of parameters from CMDs is 
fundamental. The simplest approach is to compare a given CMD with those built with 
stellar populations of previously known parameters. Depending on the quality of the CMD (basically, 
the definition of the evolutionary sequences), this exercise should yield relative values of the age, 
distance from the Sun and metallicity. A more quantitative approach is trying to find one isochrone 
(i.e., a population of stars having the same age and metallicity but different masses) that fits 
the overall CMD morphology of a star cluster. In principle, this can be - and has been - done 
{\em by eye}, but usually restricted to a small number of isochrones. This method is not practical 
for large sets of isochrones.

However, a deeper interpretation of CMDs in terms of intrinsic cluster properties such as mass, 
age, metallicity, binary fractions, etc, requires more than estimates or relative quantities. Thus, 
several approaches have been built for extracting reliable parameters from CMDs of star clusters. For 
instance, a Bayesian technique to invert CMDs of main-sequence and white dwarf stars has been introduced
by \citet{vHip06} and further developed by \citet{DeGen09} and \citet{vDyk09}. \citet{NJ06} apply 
maximum-likelihood statistics to Hess diagram\footnote{Hess diagrams (\citealt{Hess24}) contain the 
relative density of occurrence of stars in different colour-magnitude cells of the Hertzsprung-Russell 
diagram.} simulations (including binaries) to derive distances from the Sun and ages. However, they do 
not consider differential reddening, and their method appears to be more efficient for clusters older 
than $\sim30\,\my$. \citet{Hill08} model CMDs of star-forming regions and young star clusters by means 
of varying star-formation histories. \citet{daRio2010} include differential reddening, age spreads, 
and pre-main sequence stars to the \citet{NJ06} method, but have to adopt distance from the Sun and 
reddening values from previous works. \citet{LK2002} use likelihood statistics to identify the synthetic 
CMD that best reproduces the observed one, with membership probabilities determined from comparison 
among CMDs extracted from the cluster and a control field. More recently, \citet{simASA} simulated CMDs
including cluster (stellar) mass, age, distance modulus, star-formation spread, foreground and differential 
reddening, and binary fraction to reproduce the observed CMDs of young star clusters.

In general, the above approaches are designed to provide a thorough description of the stellar 
population properties giving rise to an observed CMD. Unfortunately, they usually invoke many
free parameters and assumptions; besides, they also tend
to overlook some degree of degeneracy associated with, for instance, differential reddening, age 
spread and binaries. They also do not account for the fact that the low photometric completeness 
affecting mainly the faint (low-mass) stars - especially at the central region of populous star 
clusters - may mimic a flat mass function, thus leading to confusion between actual dynamical evolution 
and completeness effects.

In many - if not most - cases, the interest is just to extract reliable values for the age, 
metallicity, reddening and distance from the Sun of a star cluster, based only on the observed CMD 
and a set of isochrones. That's the perspective we'll explore in this work.
   
This paper is organised as follows: in Sect.~\ref{approach} we describe the new isochrone fitting approach. 
In Sect.~\ref{simCMDs} we test the parameter recovery efficiency with simulated CMDs. In Sect.~\ref{realSC} 
we apply the approach to actual star clusters and discuss the results. Concluding remarks are given in 
Sect.~\ref{Conclu}.

\section{Fitting the Hess diagram}
\label{approach}

The first step of the approach consists in building the Hess diagram from the photometric measurements 
of the observed stars in a given star cluster. As a compromise between resolution and computational 
time, the Hess diagrams used in this work are built with magnitude and colour cells of dimension 
$\Delta\,mag=0.02$ and $\Delta\,col=0.01$, respectively. The smearing effect introduced by photometric 
uncertainties is explicitly considered in the Hess diagrams. Assuming that uncertainties follow a normal 
distribution around the mean values, we compute the magnitude and colour fraction (summing for all stars) 
that occurs in a given cell of the Hess diagram. In practice, this value corresponds to the difference 
of the error functions computed at the cell borders. Perhaps, this rather continuous nature is the most 
important advantage of Hess diagrams over the discreteness of CMDs. By definition, the sum of the colour 
and magnitude density over all Hess cells is the number of input stars.

In short, the present approach consists in finding the isochrone (or range of isochrones) that best 
maps the high-density ridge occurring along the evolutionary sequences. Quantitatively, this
refers to the isochrone that crosses through the largest possible number of high-density cells in 
the Hess diagram. This measure is quantified by means of the fit index (\fI), which contains the total 
Hess-density sum of a given isochrone. Specifically, for an isochrone (characterized by a given age and 
metallicity) displaced on the Hess diagram by some amount of distance modulus and colour-excess, the 
fit index is quantified as
\begin{equation}
\label{fI}
\fI = \sum_{i=1}^N H(m_i,c_i), 
\end{equation}
where $N$ is the number of magnitude points in the isochrone,
$m_i$ and $c_i$ are the magnitude and colour corresponding to the $i$th isochrone point, and 
$H(m_i,c_i)$ is the respective Hess density. Additionally, this procedure naturally leads to lower 
values of \fI\ for secondary sequences (i.e. lower-Hess density) occurring in, e.g. high binary fractions.

\begin{figure}
\resizebox{\hsize}{!}{\includegraphics{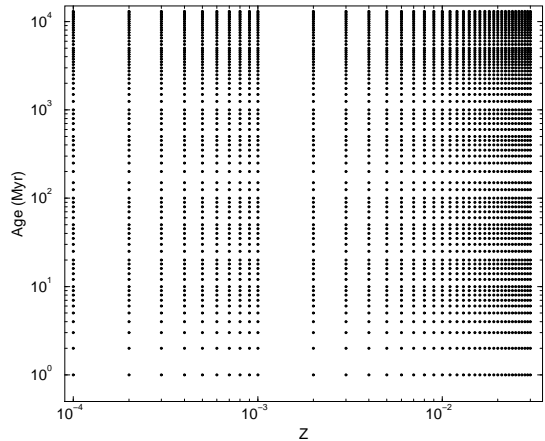}}
\caption{Each point represents an available isochrone built with the corresponding metallicity 
and age.}
\label{fig1}
\end{figure}

Given the wide-range coverage in age, metallicity, and mass - and the relative simplicity in obtaining 
them - we work here with the PARSEC isochrone set (\citealt{PARSEC}), version 1.2S\footnote{available 
at {\em http://stev.oapd.inaf.it/cgi-bin/cmd}.}. These isochrones are computed for a scaled-solar 
composition and follow the relation $Y=0.2485+1.78Z$. They consider the solar metal content as 
$\zs=0.0152$. It should be noted that our approach works with any isochrone set, as long as the 
isochrones cover the ranges of age, metallicity and stellar mass expected to occur in star clusters. 

Regarding ages and the total metal content $Z$, we consider isochrones distributed within 
$0.0001\le Z\le0.03$ and ages in the range $1\,\my\le t_A\le13\,\gy$, with higher resolution both 
for lower metallicity and younger ages. In summary, we work here with a base of 2808 isochrones. For 
clarity, the distribution of isochrones on the $Z\times$Age plane is also shown in Fig.~\ref{fig1}. 
Obviously, the approach can be equally used with the full or restricted ranges of metallicity and 
age.

Once the metallicity and age ranges are selected, each isochrone belonging to those ranges is used 
to search for the values of apparent distance modulus (DM) and colour excess (CE) that produce the 
best fit to the Hess diagram. In the present context, a fit represents the density sum over all 
Hess cells crossed by the isochrone (see eq.~\ref{fI}). The search ranges for DM and CE must also be 
given at the start. Thus, our procedure reduces to two the dimensionality of the problem, i.e.,
finding the best values of DM and CE. This might lead to losing some information on the error 
propagation in both parameters. However, as discussed in Sect.~\ref{simCMDs}, the final errors in DM 
and CE (and the remaining parameters) are computed based on the morphology of the plane Z$\times$Age 
as a function of \fI. The search for the best values of DM and CE is carried out by the global 
optimisation method known as adaptive simulated annealing (ASA), which is relatively time efficient 
and robust (e.g. \citealt{ASA}). ASA is able to distinguish and escape from different local maxima in 
its search for the absolute maximum (e.g. \citealt{simASA}). 

As a technical caveat, we remark that an isochrone is a discrete set of stellar mass, magnitudes and 
colour values. But, by construction, the number - and consequently, distribution - of these points is 
not the same for all isochrones. So, isochrones having a higher number of points would naturally end 
up being favoured by the present approach. To avoid this, each isochrone is regularised at the start, 
so that the line connecting 2 consecutive isochrone points is discretised with the same spacing as that 
adopted for the Hess cells; additionally, each cell crossed by an isochrone is counted a single time. 

The end result of applying the above procedure to all isochrones (belonging to the selected age and 
metallicity range) is the solution map, which contains the fit index for all values of $Z$ and age. For 
visualisation purposes, this map is first normalised by the \fI\ corresponding to the best solution. 
Then, the map resolution is increased by means of a bi-cubic spline interpolation. The latter step is 
necessary to find the significant solutions and compute their uncertainties (Sect.~\ref{simCMDs}). In
the context of the present work, a solution would correspond to a region (hot-spot) in the map where 
the highest values of \fI\ are concentrated. Then, the optimum values of age, metallicity, apparent 
distance modulus, and colour excess (and, consequently, distance from the Sun) can be derived from the 
hot-spot morphology. 

\begin{figure}
\resizebox{\hsize}{!}{\includegraphics{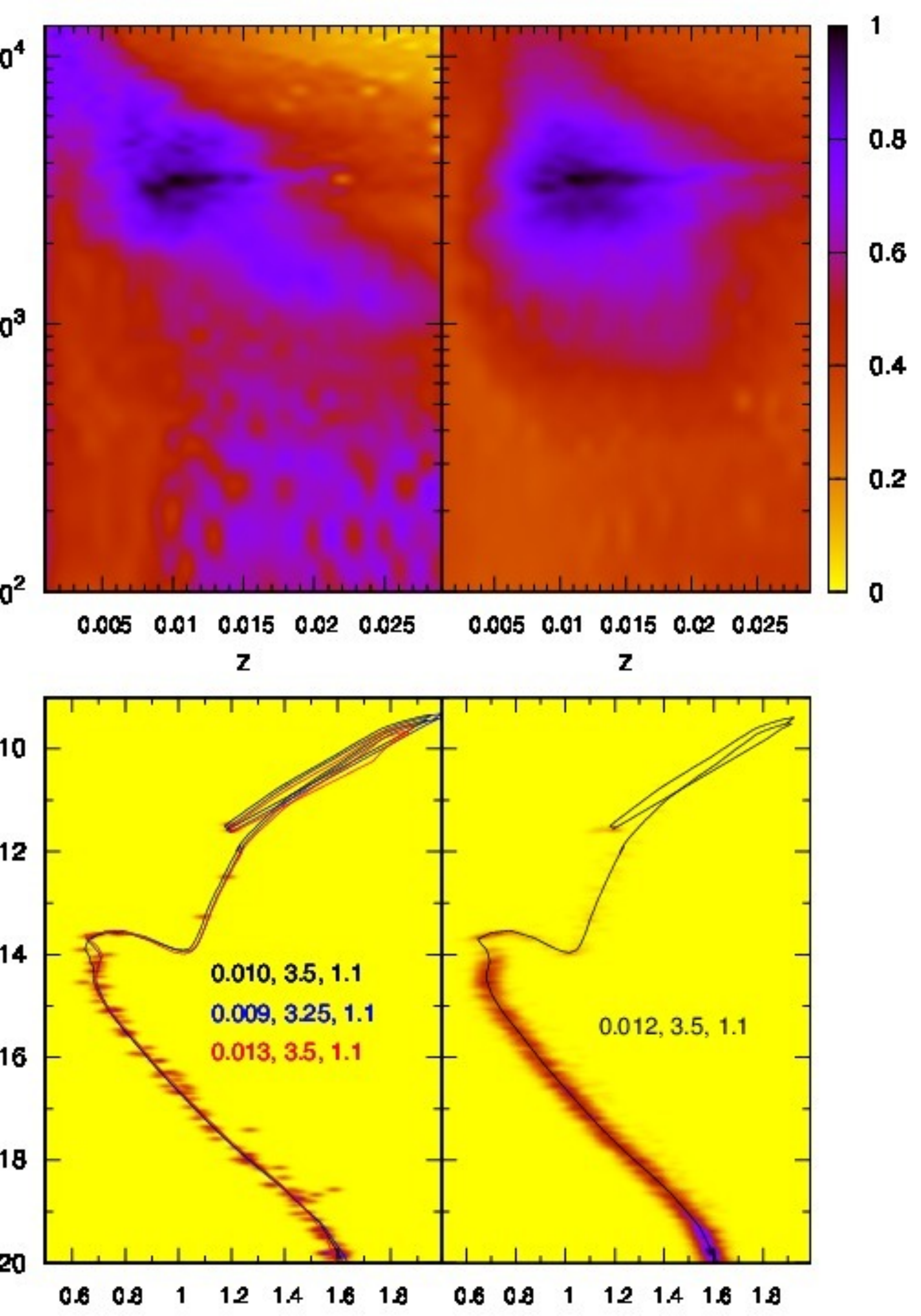}}
\caption[]{Top: $Z\times$Age solution maps for Models\#1 and \#2. The fit index scale is shown 
by the colour box (top-right). Bottom: $V\times(B-V)$ Hess diagrams together with the best-solution isochrones. 
Isochrone legends show the metallicity (Z), age (Gyr) and distance from the Sun (kpc).}
\label{fig2}
\end{figure}

\section{Testing on simulations}
\label{simCMDs}

In this section we employ simulated CMDs to investigate the efficiency of our approach in 
retrieving the input values of relevant astrophysical parameters under very different numbers
of stars. The model CMDs correspond to relatively old star clusters with total stellar masses 
of 250\,\ms (Model\#1) and 10000\,\ms (Model\#2), located at 1.1\,kpc from the Sun. For the sake 
of realism, we assume that at this distance, B and V photometry obtained with typical ground-based
telescopes would reach apparent magnitudes $\sim20$ still with relatively small errors. 

For consistency with the remainder of this work, photometric errors at given B and V magnitudes 
are simulated following the average error distribution as a function of magnitude observed for 
the star clusters M\,67, NGC\,6791 and NGC\,2635 (Sect.~ \ref{realSC}). Formally, for stars fainter
than $B=10$ and $V=10$, the corresponding photometric uncertainties are computed according to 
$\sigma_B = 0.01 + 1.0\times10^{-13}e^{(B/1.1)}$ and $\sigma_V = 0.01 + 1.5\times10^{-13}e^{(B/1.2)}$; 
brighter stars have $\sigma_B = 0.01$ and $\sigma_V = 0.01$. The simulated stellar mass 
distribution follows a \citet{Kroupa01} mass function, with each star being attributed B and V 
magnitude values according to its mass. The mass/light relation was extracted from a specific 
PARSEC isochrone built according to the Bessell \& Brett (\citealt{Bessell90}; \citealt{BesBret88}) 
photometric system. 

Except for the total stellar mass, the models are identically built with the isochrone corresponding
to 3.5\,\gy of age and total metallicity $Z=0.012$; additional model parameters are the colour 
excess $\EBV=0.21$, and the binary fraction $\fB=0.3$. At 3.5\,\gy, the corresponding isochrone
contains stars in the mass range $0.09\le m(\ms)\le1.4$. For more realism, photometric errors and 
scatter are also added to the stars. Models\#1 and \#2 end up having $\sim100$ and $\sim5000$ stars
brighter than $V=20$, respectively. The difference in number of stars, as well as the restricted
photometric range, are useful for testing the efficiency of the approach under different morphology 
definition of the evolutionary sequences on the corresponding Hess diagrams. 

\begin{figure}
\resizebox{\hsize}{!}{\includegraphics{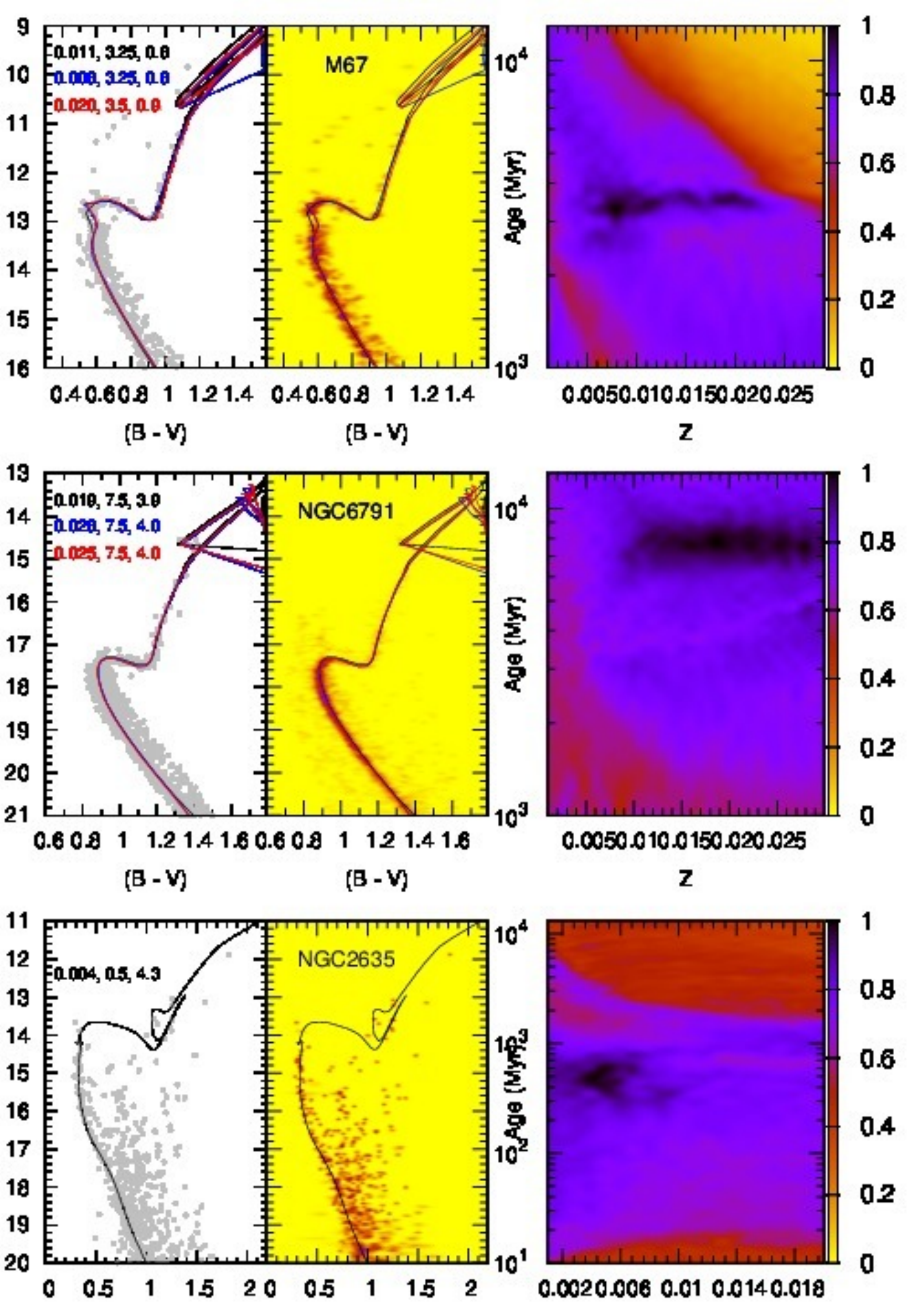}}
\caption[]{Analysis of M\,67 (top panels) and respective model (bottom). Age and metallicity 
values may differ from those in Table~\ref{tab1} because they correspond to those of the 
nearest-available isochrones. The fit index scale is shown by the colour boxes (right). 
Isochrone legends show the metallicity (Z), age (Gyr) and distance from the Sun (kpc).}
\label{fig3}
\end{figure}

To test the approach we considered all isochrones having age and metallicity within the ranges 
$100\,\my - 13\,gy$ and $0.001 - 0.030$, respectively. Fig.~\ref{fig2} (top panels) shows the 
normalised, high-resolution solution maps for the three models. Clearly, the convergence pattern 
on the age$\times$metallicity plane gets tighter as the number of stars (and so, the definition
of the evolutionary sequences on the Hess diagram) increases, which should be reflected both on 
the number of secondary solutions and the scale of the fit parameter uncertainties (see below). 
As expected, the best solution corresponds to the point with the highest \fI\ in the map. After 
locating it, we fit the map with a two-dimensional Gaussian, with axes along the $Z$ and age 
directions; the values of age and $Z$ at the maximum \fI\ are kept fixed. The free parameters, 
the Gaussian amplitude together with $Z$ and age dispersions ($\sigma_Z$ and $\sigma_A$) are 
searched again with ASA. The best-fit values of $\sigma_Z$ and $\sigma_A$ are taken as the global 
uncertainties of $Z$ and age. This Gaussian is subsequently subtracted from the map, and the new 
(if any) maximum is searched for, thus giving rise to a set of solutions ranked by the respective 
\fI. Each solution corresponds to a single value for DM and CE. Then, considering all solutions 
occurring within $\pm\sigma_Z$ and $\pm\sigma_A$, we compute the weighted (using \fI\ as weight) 
average and $1\sigma$ dispersion for both DM and CE. As a caveat we remark that, although this 
approach has the advantage of allowing the detection and characterization of multiple solutions, 
the \fI\ morphology around the solutions in Figs.~\ref{fig2} and \ref{fig3} clearly deviate from
a simple Gaussianity. The extent to which the deviation affects the parameter uncertainties and
alternative approaches for finding errors will be explored in forthcoming work. The simulated CMDs 
and respective Hess diagrams are also shown in Fig.~\ref{fig2} (middle and bottom panels), together 
with the isochrones and parameters corresponding to the best solutions. Since the optimum $Z$ and 
age are searched for on the interpolated, high-resolution map, the numerical values returned do not 
necessarily coincide with those of the input isochrones. 

The best-fit parameters obtained in the simulations are given in Table~\ref{tab1}, where we 
restrict to solutions having $\fI>0.7$. Reflecting the low number of stars and, thus, less 
constrained CMD (and Hess diagram) morphology, Model\#1 has 3 solutions with $\fI>0.7$. But, 
they all are consistent with the input values, within the uncertainties.

\begin{table}
\caption[]{Parameters for models and star clusters}
\label{tab1}
\renewcommand{\tabcolsep}{1.8mm}
\begin{tabular}{ccccc}
\hline\hline
\fI& Age & $Z$ & \EBV & $d_\odot$ \\
         &(\my)&   & (mag)&(Kpc)\\
(1) & (2) & (3) & (4) & (5) \\
\hline
\multicolumn{5}{c}{Model\#1 (109 stars)}\\
\hline
1.00 & $3462\pm160$ & $0.010\pm0.001$ & $0.23\pm0.01$ & $1.07\pm0.02$\\
0.85 & $3173\pm290$ & $0.009\pm0.001$ & $0.26\pm0.02$ & $1.07\pm0.03$\\
0.83 & $3519\pm135$ & $0.013\pm0.001$ & $0.20\pm0.02$ & $1.11\pm0.02$\\
\hline
\multicolumn{5}{c}{Model\#2 (5011 stars)}\\
\hline
1.00 & $3500\pm120$ & $0.012\pm0.001$ & $0.22\pm0.01$ & $1.08\pm0.02$\\
\hline
\hline
\multicolumn{5}{c}{M\,67 (446 stars)}\\
\hline
1.00 & $3292\pm300$ & $0.008\pm0.002$ & $0.16\pm0.03$ & $0.79\pm0.03$\\
0.82 & $3542\pm150$ & $0.019\pm0.003$ & $0.02\pm0.02$ & $0.88\pm0.02$\\
\hline
\multicolumn{5}{c}{NGC\,6791 (2092 stars)}\\
\hline
1.00 & $7694\pm505$ & $0.019\pm0.001$ & $0.23\pm0.02$ & $3.87\pm0.07$\\
0.97 & $7472\pm633$ & $0.028\pm0.002$ & $0.17\pm0.01$ & $4.03\pm0.06$\\
0.85 & $7556\pm250$ & $0.025\pm0.001$ & $0.18\pm0.02$ & $4.02\pm0.06$\\
0.78 & $8000\pm440$ & $0.012\pm0.001$ & $0.29\pm0.02$ & $3.67\pm0.07$\\
\hline
\multicolumn{5}{c}{NGC\,2635 (481 stars)}\\
\hline
1.00 & $500\pm100$ & $0.004\pm0.001$ & $0.39\pm0.02$ & $4.28\pm0.15$\\
\hline
\end{tabular}
\begin{list}{Table Notes.}
\item Col.~1: Normalised fit index; Col.~2 Age; Col.~3: Total metal content; Col.~4: Colour 
excess; Col.~5: Distance from the Sun.
\end{list}
\end{table}

In summary, tests with simulated CMDs show that our approach is efficient in recovering the 
input parameters and providing reasonable uncertainties.

\section{Testing on star clusters}
\label{realSC}

At this point we test our approach with actual star clusters. For this exercise we selected the 
relatively old open clusters M\,67 (NGC\,2682) and NGC\,6791, and the younger NGC\,2635. They
were selected for having relatively recent and uniform CCD photometry easily accessible through 
{\bf VizieR}\footnote{http://vizier.u-strasbg.fr/viz-bin/VizieR}. In addition, they are characterised
by different ages, metallicities, number of stars, distances from the Sun and amount of field-star 
contamination. Both M\,67 and NGC\,6791 are older and closer to the Sun and, perhaps, more populous 
than NGC\,2635. One consequence is that both older clusters produce clearly discernible evolutionary 
sequences in CMDS or Hess diagrams (Fig.~\ref{fig3}). NGC\,2635, on the other hand, has a more 
ambiguous (less populated and more contaminated) CMD morphology. Thus, these three clusters provide
some variety of conditions for testing our approach.

\subsection{M\,67}
\label{m67}

The high Galactic latitude of M\,67 ($b=+31.9\degr$) is an advantage for minimising field-star 
contamination, which makes it a convenient cluster for this analysis. Several works have come up
with a nearly solar metallicity for M\,67 (e.g. \citealt{Sarajedini09}). The 
{\bf WEBDA}\footnote{http://www.univie.ac.at/webda/} database provides $\ds=0.91$\,kpc, $\EBV=0.06$, 
and the 2.6\,\gy of age. More recently, the study of \citet{Chen14} assumed $Z=0.02$, $\mMo=9.75$, 
and $\EBV=0.03$, with which they derived the significantly older value of 3.5\,\gy as the the most 
probable age of M\,67. 

Our CMD of M\,67 consists of B and V photometry taken from the ground-based CCD astrometry catalog 
of \citet{Yadav08}. We selected stars lying closer than 20\,\arcmin\ from the cluster center, thus 
resulting in 446 stars brighter than $V=16$. As expected, the evolutionary sequence of a relatively
old open cluster stands out clearly in the CMD and Hess diagram, with a minimum field contamination
(Fig.~\ref{fig3}).

Solutions have been searched for isochrones having age within $1 - 13\,\gy$ and metallicity within 
$0.001 - 0.03$. Our approach finds 2 solutions having $\fI>0.7$ for M\,67. The results are given 
in Table~\ref{tab1}, and the corresponding solution maps, Hess diagrams and CMDs are shown in 
Fig.~\ref{fig3}. The solution map pattern shows a rather tight distribution of ages around 
$\sim3.5\,\gy$, but the metallicity distributes over the relatively wide range $0.06\la Z\la0.22$.
Within the uncertainty, the age we find for M\,67 - corresponding to the best fit - is nearly the 
same as that found by \citet{Chen14}, but our metallicity is about half theirs. The second-ranked 
solution is rather consistent in age, but presents an increase in metallicity. Interestingly, the 
isochrone fit adopted by \citet{Chen14} is recovered by our approach as the second-ranked solution, 
with $\fI=0.82$. For comparison purposes, Fig.~\ref{fig3} shows the two solutions superimposed to 
the CMD and Hess diagram of  M\,67. 

\subsection{NGC\,6791}
\label{n6791}

NGC\,6791 is closer to the Galactic plane ($b=+10.9\degr$) and somewhat older and metal richer than 
M\,67. For instance, WEBDA gives $\ds=4.1$\,kpc, $\EBV=0.12$, 4.4\,\gy, and $Z=0.21$ for NGC\,6791. 
More recently, \citet{vdB14} (and references therein) use $[Fe/H]\sim0.3$ (thus implying $Z\sim2\zs$) 
and derive an age of $\sim8\,\gy$ for NGC\,6791.
  
The B and V photometry for NGC\,6791 was taken from the CCD catalogue of \citet{Stet03}. To minimise
field-star contamination, we only considered stars closer than 3\,\arcmin\ from the center and brighter 
than $V=21$, resulting in $\sim2100$ stars. Even so, the CMD and Hess diagram of NGC\,6791 clearly
contain some degree of field contamination (Fig.~\ref{fig3}). 

Solutions were searched within the same age and metallicity ranges as for M\,67. We find 4 solutions 
having $\fI>0.7$ for NGC\,6791, and they all consistently confirm its older age with respect to M\,67,
probably reaching 8\,\gy. Regarding the metallicity, the highest-ranked solutions indicate 
$0.019<Z<0.028$, but with a value as low as $Z=0.012$ with a low probability. Thus, the metallicity 
appears to be higher than solar, probably reaching nearly twice the solar value (Table~ \ref{tab1}). 
These features (i.e. the relatively narrow age range associated with some dispersion in metallicity) 
are present in the solution map. 

The three best-fit PARSEC isochrones produce an excellent description of the entire evolutionary 
sequence of NGC\,6791. However, the difference in \fI\ among the first three 
solutions is rather small, which means that they are statistically significant as well. The weighted
average of the 4 solutions (using the respective \fI\ as weight) yields the age $7668\pm193\,\my$,
$Z=0.021\pm0.006$, $\EBV=0.21\pm0.05$, and $\ds=3.90\pm0.14$\,kpc. An age between 7.5\,\gy and 8\,\gy 
(in our case derived for most of the main sequence and RGB$+$AGB stars) is consistent with the value 
obtained by \citet{GB10} based on properties of the white dwarf cooling sequence of NGC\,6791.

\subsection{NGC\,2635}
\label{n2635}

Located in the third Galactic quadrant and at $b=+3.96\degr$, the field of NGC\,2635 is naturally more
contaminated than both of M\,67 and NGC\,6791. Combined with a large distance from the Sun and a small
population, the available CMDs of NGC\,2635 lead to somewhat discrepant values of its fundamental 
parameters. For instance, WEBDA provides $\sim300\,\my$ of age, $\ds=5.7$\,kpc, and $\EBV=0.35$. 
\citet{Moi06} employ CCD broad band photometry and CMDs to find $\sim600$\,Myr of age $\ds\sim4$\,kpc, 
$\EBV=0.35$, and the very-low metallicity $Z=0.004$ for NGC\,2635. 

Photometry for NGC\,2635 was taken from the CCD broad band UBV catalogue of \citet{Moi06}, via
{\bf VizieR}. Because of the previous evidence of a younger age and lower metallicity (with respect 
to M\,67 and NGC\,6791), solutions have been searched for isochrones with age within $10\,\my- 13\,\gy$
and metallicity within $0.001 - 0.02$.

The CMD and Hess diagram of NGC\,2635 clearly reflect its small stellar content (e.g. the rather
discrete distribution of stars along the main sequence) and unaccounted-for field-star contamination.
Nevertheless, our approach indicates $\sim500$\,Myr as the most probable age for NGC\,2635, together with 
the low metallicity of $Z=0.004$. Interestingly, this age corresponds to the average between those of 
WEBDA and \citealt{Moi06}). Metallicity, reddening, and distance from the Sun agree with those of 
\citet{Moi06}. The best-fit isochrone passes through the main sequence and a group of relatively red and 
bright stars that - in this case - would correspond to giants (Fig.~\ref{fig3}). It's obvious that, if 
some field-star contamination is present among the giants, the end result of the age may change, but still 
probably within the $100\,\my$ uncertainty derived by our approach. Nevertheless, the \fI\ map shows that 
age and metallicity are quite concentrated around $\sim500\,\my$ and $Z=0.004$.

Finally, we note that the fit index patterns in the solution maps of M\,67 and NGC\,6791 present 
a high degree of similarity. Basically, they are relatively constrained in age but somewhat wide 
in metallicity, which explains the $Z$-distribution among the solutions for M\,67 and NGC\,6791
(Table~\ref{tab1}). The apparent degeneracy in metallicity associated with isochrones - especially 
the old ones - is a clear example of the difficulties related to deriving precise values of $Z$ 
from photometry. Ideally, isochrones covering a wider range of wavelengths - especially towards
the blue/violet - should mitigate this degeneracy and, as discussed in previous sections, our approach 
is designed to work with any isochrone set. However, isochrones built with blue filters usually do
not yet have the same accuracy as those in B and V in describing observed evolutionary sequences in
CMDs. This occurs especially because of the lack of a robust determination of molecular opacities, 
incorrect bolometric corrections and colour-$T_{eff}$ relations, and incomplete convection description, 
mostly related to low-$T_{eff}$. This issue is explored and highlighted, for instance, in \citet{CKBD13}, 
and references therein.

\section{Summary and conclusions}
\label{Conclu}

In this paper we describe a relatively simple and direct approach designed to extract 
astrophysical parameters (specifically the age, metallicity, foreground reddening and 
distance from the Sun) from CMDs of star clusters. Central to the approach is the 
availability of isochrone databases covering wide ranges - and with a high resolution - in 
age and metallicity. For this reason, we work here with the PARSEC (\citealt{PARSEC}) isochrone 
database, but any comprehensive (in terms of age and metallicity) isochrone set should work
as well. 

The basic idea is to search for the isochrone that best maps the high-density regions 
of the Hess diagram corresponding to the input CMD. This is accomplished by varying the 
apparent distance modulus and reddening for each isochrone. At each step, the fit index
(\fI) corresponding to the sum of the Hess density spanned by the isochrone is computed. 
The best values of distance modulus and reddening are those that lead to the highest 
possible \fI\ for each isochrone. Finally, a map of \fI\ is produced on the 
age$\times$metallicity plane, in which the solutions and uncertainties are searched for.

Tests with simulated CMDs show that the approach is efficient in recovering the input
values. Obviously, convergence (and uncertainties) depends somewhat on the Hess diagram
morphology. Star cluster CMDs containing large number of member stars tend to produce 
well-defined evolutionary sequences in Hess diagrams. This, in turn, increases convergence 
of the approach and decreases the uncertainties in the retrieved parameters.

From a theoretical perspective, it's clear that our approach depends critically on the 
accuracy of the isochrones in describing the stellar evolutionary sequences at a given age 
and metallicity. Right now this is a minor issue, because several groups are actively working 
on the development of isochrones based on state-of-the-art stellar physics.

\section*{Acknowledgements}
We thank an anonymous referee for important comments and suggestions.
Partial financial support for this research comes from CNPq (Brazil). This research has made use 
of the VizieR catalogue access tool, CDS, Strasbourg, France. This research has made use of the 
WEBDA database, operated at the Department of Theoretical Physics and Astrophysics of the Masaryk 
University


\end{document}